\documentclass[usenatbib, usegraphicx]{mn2e}
\bibpunct{(}{)}{;}{a}{}{,}
\def\cm2{cm$^{-2}$}
\def\hi{H\,{\sc i}}
\def\di{D\,{\sc i}}
\def\si3{Si\,{\sc iii}}
\def\mg2{Mg\,{\sc ii}}
\def\c4{C\,{\sc iv}}
\def\nv{N\,{\sc v}}

\newcommand{\lya}{Ly$\alpha$}
\newcommand{\lyb}{Ly$\beta$}
\newcommand{\kms}{~kms$^{-1}$}

\title{The Deuterium Abundance in the {\em z} = 0.7 absorber towards
QSO PG1718+4807} \author[] {N. H. M. Crighton, J. K. Webb,
R. F. Carswell and K. M. Lanzetta} \date{}

\pagerange{\pageref{firstpage}--\pageref{lastpage}}
\pubyear{}

\begin{document}

\maketitle

\label{firstpage}

\begin{abstract}
We report a further analysis of the ratio of deuterium to hydrogen
(D/H) using HST spectra of the $z=0.701$ Lyman limit system towards
the QSO PG1718$+$481.  Initial analyses of this absorber found it gave
a high D/H value, $1.8 - 3.1 \times 10^{-4}$ \citep{Webb98a},
inconsistent with several higher redshift measurements.  It is thus
important to critically examine this measurement.  By analysing the
velocity widths of the \di, \hi\ and metal lines present in this
system, \citet{Kirkman01} report that the additional absorption in the
blue wing of the \lya\ line can not be \di, with a confidence level of
98\%.  Here we present a more detailed analysis, taking into account
possible wavelength shifts between the three sets of HST spectra used
in the analysis.  We find that the constraints on this system are not
as strong as those claimed by \citet{Kirkman01}.  The discrepancy
between the parameters of the blue wing absorption and the parameters
expected for \di\ is marginally worse than $1\sigma$.

\citet{Tytler99} commented on the first analysis of \citet{Webb97a,
Webb98a}, reporting the presence of a contaminating lower redshift
Lyman limit system, with log[$N$(\hi)] $= 16.7$ at $z=0.602$, which
biases the $N$(\hi) estimate for the main system.  Here we show that
this absorber actually has log[$N$(\hi)] $< 14.6$ and does not impact
on the estimate of $N$(\hi) in the system of interest at $z = 0.701$.

The purpose of the present paper is to highlight important aspects of
the analysis which were not explored in previous studies, and hence
help refine the methods used in future analyses of D/H in quasar
spectra.

\end{abstract}


\section{Introduction}

Virtually all the deuterium (D) and the vast majority of H and $^4$He
we observe today were produced during big bang nucleosynthesis (BBN).
BBN also produced small amounts of $^3$He and $^7$Li.  In standard BBN
theory, the primordial abundances of all these light elements depends
solely on the cosmological baryon density, $\Omega_{b}$. This allows
$\Omega_{b}h^2$ to be measured directly by finding the primordial
ratio of any two of these light elements\footnote{$h \equiv
\frac{1}{100 ~ km~s^{-1}Mpc^{-1}}\times\rm{(Hubble\ constant)}$}.  The
ratio D/H is very sensitive to $\Omega_{b}h^2$, and of all the BBN
light elements, D/H potentially gives the best constraints on
$\Omega_{b}$.

QSO absorption systems provide a unique way to measure the primordial
abundance of deuterium \citep*{Adams76}.  Theoretically there are
processes other than BBN that create D, but it has been shown
\citep*{Epstein74} that the D production due to these processes is
likely to be negligible compared to D production during BBN (but see
also \citealt*{Jedam02}).  Stellar nuclear reactions cause a net
destruction of D, so measurements of the D abundance in sites that
have undergone star formation, such as our Galaxy, provide a lower
limit to the priordial D/H value.  However, many QSO absorption clouds
are thought to have undergone very little star formation, and we
expect D/H in these clouds to be close to the primordial value.
Unfortunately, very few absorption systems have the high column
density ($\ge 10^{17}\textrm{ absorbers per cm}^2$), simple velocity
structure and narrow linewidth that are required for the \di\ line to
be separated from the nearby \hi\ absorption \citep*{Webb91a}.

Most existing D/H measurements have been made in absorption systems
with a redshift large enough that the Lyman limit falls in the visible
($z > 2.2$). For lower redshift systems the Lyman limit falls in the
UV and must be observed from space.  However, it is desirable to look
for low redshift systems as there is a much larger sample of bright
QSOs at lower redshifts.  In addition, the number density of \lya\
forest absorbers is much lower at low redshifts compared to high
redshifts.  This means it is easier to accurately measure the
continuum level around the relevant absorption lines, and the chance
of a randomly placed \hi\ line contaminating the \di\ absorption is
smaller.

The absorption system at $z=0.701$ towards PG1718$+$4807 was first
identified as an excellent candidate for a D/H measurement from an
International Ultraviolet Explorer (IUE) spectrum covering its Lyman
limit (LL). Its LL is `grey', meaning the flux does not drop to zero
bluewards of the LL.  The drop in flux at the grey LL strongly
constrains the column density of the absorber.  The column density was
high enough for the \di\ \lya\ line to be detected, and the sharp
break at the LL suggested the velocity strucure of the absorption
complex was very simple.  \citet*{Webb97a} published the first analysis
of D/H in the $z=0.701$ system, using a GHRS spectrum covering the
\lya\ and \si3\ lines and the IUE spectrum covering the LL.
\citet{Webb97a, Webb98a} fitted the \hi\ absorption with a single cloud,
assuming the redshift of the \si3\ line was the same as the redshift
of the \hi\ absorption. They found D/H of $1.8 - 3.1 \times 10^{-4}$,
about ten times higher than D/H measured in other, higher redshift
absorption systems ($\sim 2 - 4 \times 10^{-5}$, see \citealt*{OMeara01,
Pettini01a, Levsh02a, Kirkman03}).

\citet*{Tytler99} subsequently published a further analysis of D/H in
the same absorber, adding Keck/HIRES spectra of \mg2\ lines associated
with the $z=0.701$ absorber.  Using the \mg2\ lines they considered
several models with different constraints on the \hi\ component's
redshift.  They found a larger range of D/H: $8 - 57 \times 10^{-5}$.

\begin{figure*}
\begin{minipage}{140mm}
\begin{center}
\includegraphics[height=1.0\textwidth,angle=-90]{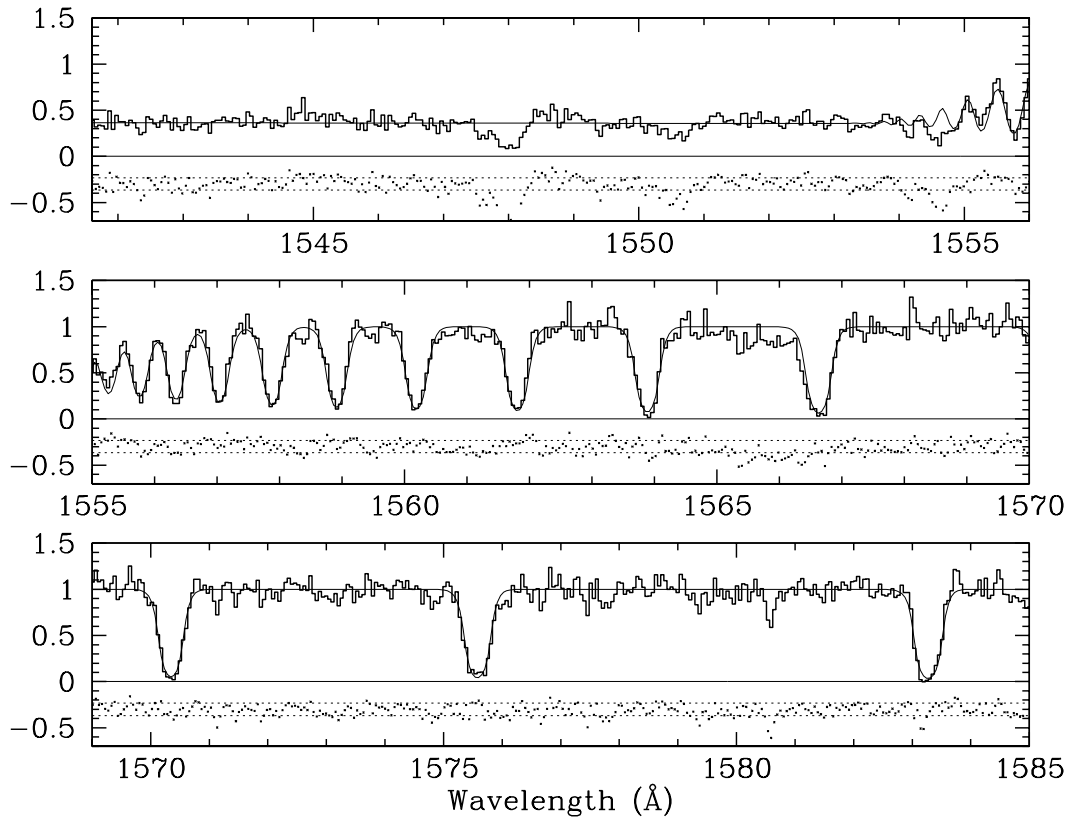}

\caption{\label{ll}The STIS spectrum of the $z=0.701$ Lyman limit, a
weighted sum of four STIS exposures with the G140M grating. There
appears to be Galactic C{\sc iv} absorption and emission at $\sim$1548
and $\sim$1550.5 \AA.  We are unsure what is causing the absorption at
$\sim$1554.5, $1565-1566.5$ and $\sim$1580.5 \AA.  Neither these
regions, nor the regions affected by Galctic C{\sc iv} absorption were
included in the fit.  The normalised residuals (which we define
$\equiv \frac{1}{15}(\textrm{data point} - \textrm{fitted
value})/\rm{error}$) and the $1 \sigma$ error levels for the residuals
are shown centred on $y=-0.3$. The thin curve shows the best fitting
solution when the \lya\ lines and the LL are fit simultaneously.  This
fit provides the \hi\ parameters in Table \ref{tab}.}
\end{center}
\end{minipage}
\end{figure*}

\citet*{Kirkman01} (hereafter KTOB) published new HST STIS
(Space Telescope Imaging Spectrograph) spectra covering the LL, \lya\
and \si3\ lines of the absorber.  The spectrum of the LL has a
resolution of $10$ \kms, and all the Lyman series lines from
Lyman-$\epsilon$ to Lyman-$16$ were resolved (see Fig.~\ref{ll}).
This spectrum confirmed the presence of a single {\em strong} \hi\
component and allowed its redshift to be measured precisely.  However,
\citet{Kirkman01} found additional {\em weak} absorption to the red of
the main \hi\ component.  This absorption was too weak to be seen in
the higher order Lyman series, but was seen in the new STIS spectrum
of the \lya\ line.  They also claimed that the $b$ parameter
($b=\sqrt{2} \sigma $, where $\sigma$ is the velocity dispersion) of
the putative \di\ was larger than that expected for a gas with a
temperature and turbulence calculated from the \hi\ and metal lines'
$b$ parameters.  Based on these two points they claimed that the
absorption fitted as \di\ in \citet{Webb97a} and \citet{Tytler99} is
contaminated by weak \hi\ absorption. Due to this contamination they
concluded the system does not provide evidence for a high D/H value.

We present a further analysis of D/H in the $z = 0.701$ absorption
system, in which we look at the effect of systematic errors, not
addressed in previous analyses, that affect the $z=0.701$ absorption
system parameters.  In particular we analyse how the the putative \di\
absorption's parameters are affected by wavelength calibration errors
in the HST spectra and present revised parameters for this particular
absorption spectrum.  These calibration effects will be important for
future analyses of D/H in QSO spectra.  Finally, we assess the
likelihood that this absorption system shows \di\ and give an estimate
of D/H for this system.

\section{Data and analysis}

\begin{figure*}
\begin{minipage}{140mm}
\begin{center}
\includegraphics[height=1.0\textwidth,angle=-90]{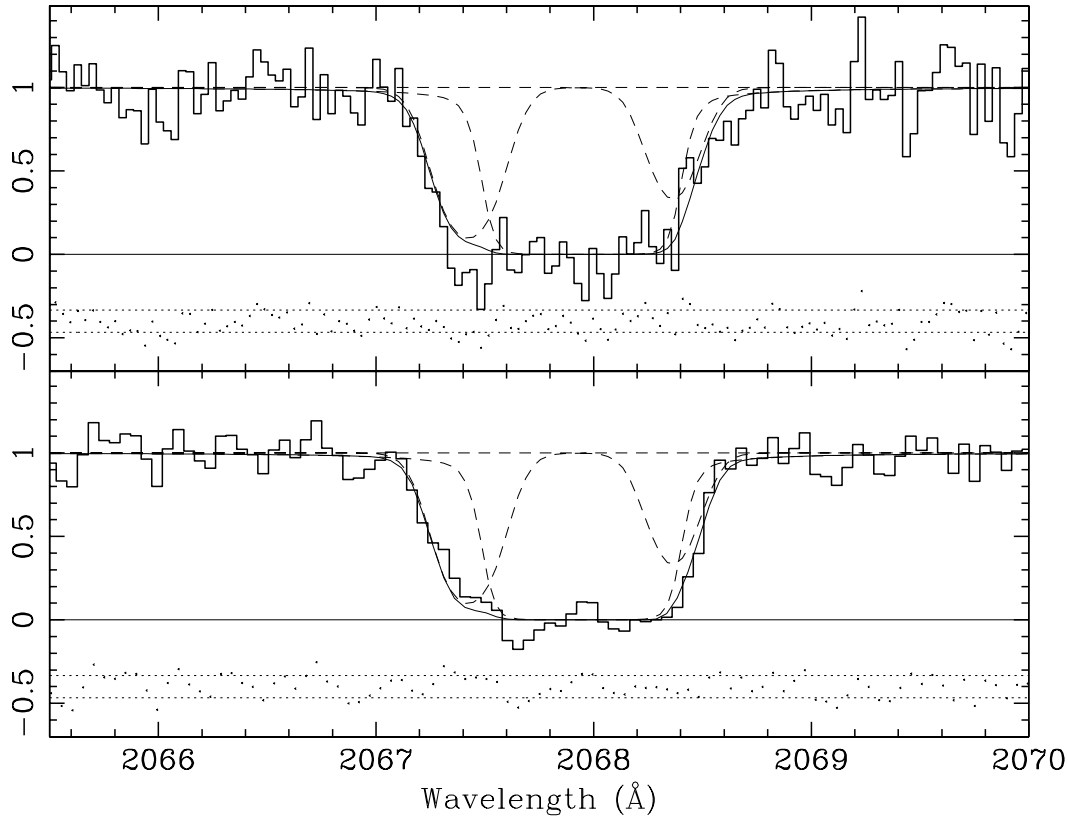}

\caption{\label{lya}The GHRS (above) and STIS (below) spectra covering
the $z=0.701$ \lya\ line. The normalised residuals and their $1
\sigma$ error levels are shown centred on $y=-0.4$.  The thin curve
shows the best fitting solution when the \lya\ lines and the Lyman
limit are fit simultaneously. This fit provides the parameters in
Table \ref{tab}. The contributions from the main \hi\ component, red
and blue components are shown by dashed curves.  Note that the `shape'
of the STIS E230M \lya\ line appears to be different to that of the
GHRS line.  In particular, the shape of the red wing and the
absorption at $\sim 2067.4$ \AA\ are different.  We are not certain
what causes these differences, but they may be explained by
correlations in the noise.}
\end{center}
\end{minipage}
\end{figure*}

\begin{figure*}
\begin{minipage}{140mm}
\begin{center}
\includegraphics[height=1.0\textwidth,angle=-90]{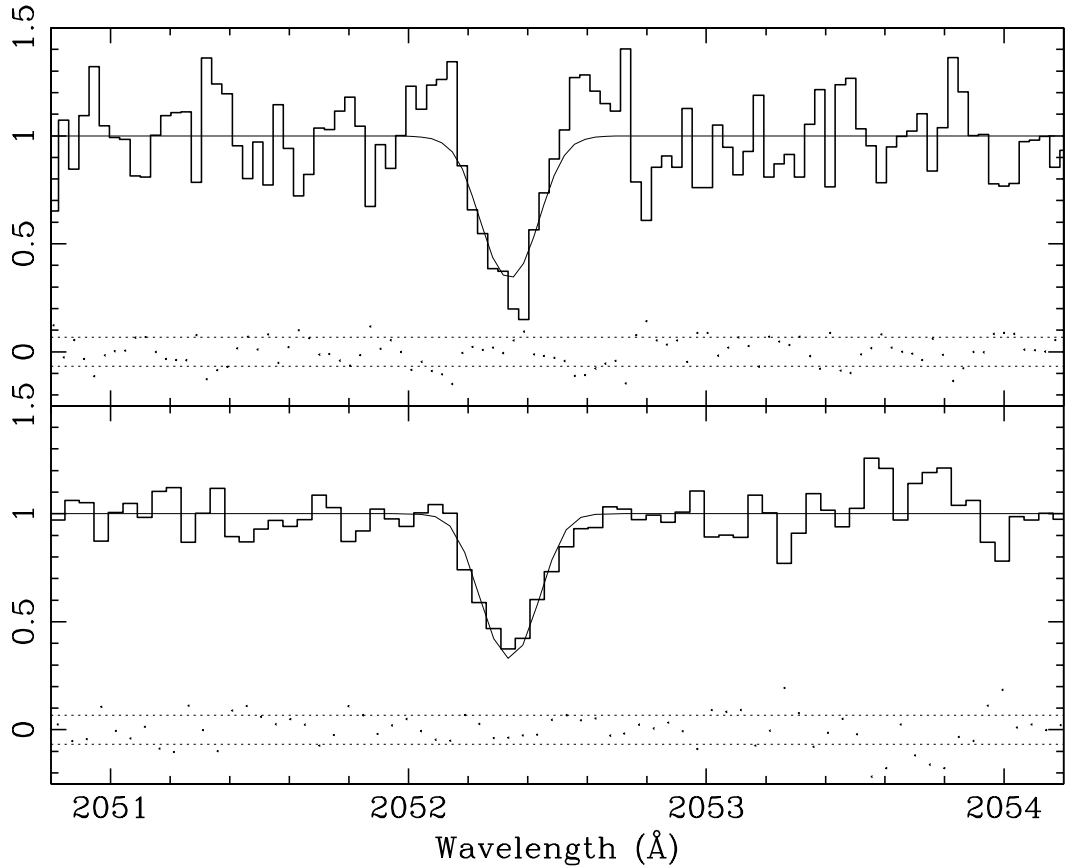}

\caption{\label{si} The GHRS (above) and STIS (below) spectra covering
the $z=0.701$ \si3\ line. The normalised residuals and their $1 \sigma$
error levels are shown centred on $y=0$. The thin curve shows the best
fitting solution when both lines are fit simultaneously.  This fit
provides the parameters in Table \ref{tab}.}

\end{center}
\end{minipage}
\end{figure*}

\begin{figure*}
\begin{minipage}{140mm}
\begin{center}
\includegraphics[height=1.0\textwidth,angle=-90]{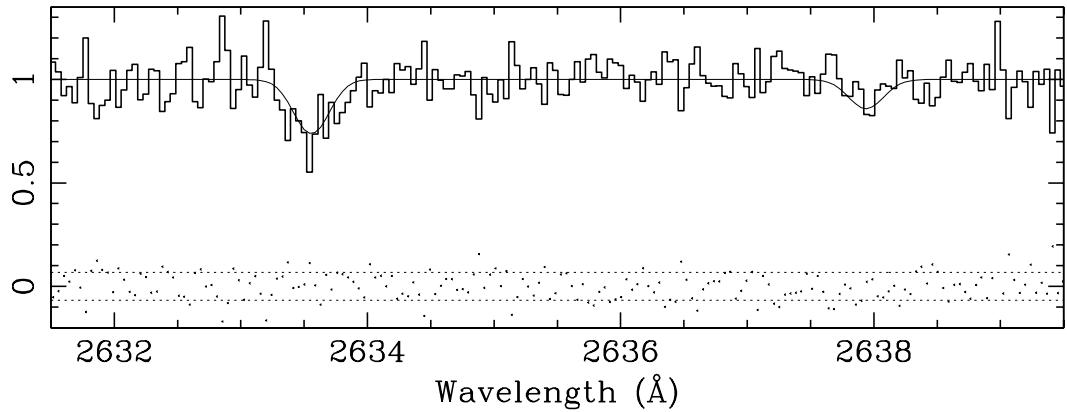}

\caption{\label{c}A tentative detection of the $z=0.701$ \c4\ doublet
in the STIS E230M spectrum. The normalised residuals and their $1
\sigma$ error levels are shown centred on $y=0$. The thin curve shows
the best fitting solution when both lines of the doublet are fit
simultaneously.  This fit provides the parameters in Table
{\ref{tab}}.}
\end{center}
\end{minipage}
\end{figure*}

The QSO PG1718$+$4807 has been observed with the Faint Object
Spectrograph (FOS), Goddard High Resolution Spectrograph (GHRS) and
STIS on the HST, and the IUE \citep*{Lanzetta93a}\footnote{The GHRS,
STIS, FOS and IUE spectra are all available from the HST multimission
data archive: http://archive.stsci.edu}.  We are interested in finding
the parameters of the \hi\ and \di\ absorption.  The resolutions of
the FOS and IUE spectra are much lower (FWHM $\sim$~150 and
$\sim$~900~\kms\ respectively) than those of the GHRS and STIS
spectra. The FOS and IUE spectra do not provide any useful constraints
on the parameters of \di\ or any \hi\ sub-components which may be
present in the $z=0.701$ absorption complex.  The parameters of the
main \hi\ component are very well determined by the STIS spectra alone
(see section \ref{voigt}). Therefore we use only the GHRS and STIS
spectra in our analysis.

\citet{Tytler99} detected \mg2\ with Keck/HIRES.  We use their best
fitting parameters for \mg2, with a modified error for the \mg2\ $b$
parameter (see section \ref{voigt}).

\subsection{GHRS data}
The GHRS spectra were taken using the G140M grating with a 0.2
$\times$ 0.2 aperture giving a line spread function (LSF) with a FWHM
of $14$\kms.  The observations consisted of 56 exposures, which were
shifted and added using the IRAF routines written for this purpose by
the Space Telescope Science Institute (STScI), {\em poffsets} and {\em
specalign}.  We used the pipeline reduced spectra, processed with
CALHRS version 1.3.13. The wavelength range covered is $2049 - 2098$
\AA. The S/N per pixel near the \lya\ and \si3\ lines in the combined
spectrum is $\sim$10.

\subsection{STIS data}

The STIS observations were taken with the E230M echelle grating and
the G140M longslit grating with the 0.2$\times$0.2 aperture. The FWHM
of each of the LSFs of the STIS observations are 2 pixels for the
E230M echelle grating (5 \kms\ per pixel) and 1.5 pixels for the G140M
grating (0.06 \AA\ or $\sim$11 \kms\ per pixel).  Four exposures were
taken with the G140M grating, and five with the E230M grating.  In
each case we used the pipeline reduced spectra (CALSTIS version 2.8).
For each set of observations the exposures were rebinned to a common
wavelength scale and combined using a variance weighted average. The
E230M spectra cover the wavelength region $1842 - 2673$ \AA.  The
final S/N per pixel near the \lya\ and \si3\ lines in the averaged
E230M spectrum is $\sim$5.  The G140M spectra cover the
Lyman-$\epsilon$ transition down to the Lyman limit, a wavelength
range of $1540 - 1594$ \AA.  The S/N per pixel in the G140M averaged
spectrum is $\sim$9.

\subsection{Line spread functions}

The LSF of the GHRS observations can be approximated by a Gaussian,
but the STIS LSFs are significantly different from a Gaussian. We used
the instrumental profiles for the E230M and G140M gratings given in
the STIS instrument
handbook\footnote{http://www.stsci.edu/hst/stis/performance/spectral\_resolution/}.
The actual STIS LSFs may be slightly asymmetric. The nature of this
asymmetry depends on the orientation of the slit on the sky (Sahu in
KTOB).  We averaged the asymmetric sides of the LSF to
give a symmetric LSF for both the G140M and E230M gratings. We note
that even if some small asymmetry is present, it will not have a
significant effect on the fitted absorption line parameters.  Once the
LSF per pixel for the original spectrum was determined it was rebinned
to the same wavelength scale as our final averaged spectrum.  For the
E230M grating, the pixels are a constant size in velocity rather than
wavelength.  We calculated an E230M LSF at the central wavelength of
each line fitted in the E230M spectrum.

\subsection{Continuum placement}

The placement of the continuum in the STIS G140M spectrum has a
potentially significant effect on the fitted $b$ parameter and column
density of the main \hi\ component.  To test the magnitude of this
effect, we fitted the continuum of this region using several methods.
Firstly we fitted a constant flux to regions apparently free of
absorption. Secondly we fitted a power law $f_{\lambda} \propto
\lambda^{-\alpha}$.  Finally, the IUE spectrum covers a larger
wavelength range redwards of the Lyman limit than the STIS G140M
spectrum, so it may give a better estimate of the continuum than just
fitting the STIS Lyman limit.  We fitted the IUE Lyman limit with a
3rd order Chebyshev polynomial and scaled this continuum to the level
of the STIS Lyman limit, based on regions apparently free from
absorption. Each of these methods gave very similar column densities
and $b$ parameters for the main \hi\ component. The small error due to
continuum fitting was added in quadrature to the statistical error.
This combined error is the first error given in Table \ref{tab}.

The error in continuum placement around the \lya\ line, \si3\ line and
\c4\ doublet (Fig.~\ref{lya}, \ref{si} and \ref{c}) was determined by
fitting both straight lines and 3rd order Chebyshev polynomials to
regions apparently free from absorption close to each line.  The
differences between these two methods of continuum fitting have a
negligible effect on the absorption line parameters.

\subsection{Wavelength calibration and shifts between the HST spectra}
\label{shift}
Misalignment of the wavelength scale between the GHRS, STIS G140M and
STIS E230M spectra can potentially affect all the fitted parameters
for \hi\ and the putative \di, and the redshift for \si3 and \c4.

The GHRS wavelength scale appears to be offset from the STIS
wavelength scale by $\sim 0.07$ \AA\ ($\sim 1.4$ GHRS pixels).
Despite helpful discussion with Claus Leitherer from STScI we were
unable to discover the reason for the discrepancy.  KTOB
shifted the GHRS spectrum to coincide with the STIS spectrum, finding
the amount by which to shift by cross correlation over the common
wavelengths covered. They then checked this shift by comparing the
positions of profile fits to sharp features in each spectrum. However,
they do not estimate the error in their shift.

We calculate the shift by choosing the three sharpest absorption
features at relatively high S/N present in each spectrum and
calculating the cross correlation for each of them. The regions we
used were: $2051.1-2053.7$ \AA\ (the \si3\ line), $2056.5-2060.4$ \AA\
(a \lyb\ line at $z=1.0065$) and $2069.6-2072.2$ \AA\ (a \lya\ line at
$z=0.70326$).  The three shifts for these regions were 0.070 \AA,
0.072 \AA\ and 0.058 \AA\ respectively.  The three shifts are roughly
consistent, suggesting that the wavelength offset is the same across
the spectrum. We take the average of these three shifts and move the
GHRS spectrum by this amount to align it with the STIS spectrum,
assuming that the STIS wavelength calibration is more accurate than
the the GHRS calibration. To quantify the error in this shift, we fit
the sharpest feature in each spectrum, the \si3\ line, with VPFIT.
The VPFIT errors in the position of each line were then added in
quadrature to give an estimate of the total error in the shift.  This
error is $\sim 0.007$ \AA, or $\sim 1$\kms.  This error does not a
have a significant effect on the the parameters of the putative \di\
or main \hi\ absorption.

For a 0.2$\times$0.2 aperture and the MAMA detector used for the STIS
observations, the absolute wavelength calibration error can be as much
as 0.5 -- 1.0 pixels ($2 \sigma$, \citealp*{STISinstr02}).  In our
analysis we introduce a conservative error of 0.3 pixels in the G140M
and E230M wavelength calibrations.  Thus we consider three different
cases: (1) The wavelength calibration of both the G140M and E230M
spectra are both correct, (2) the E230M spectrum wavelength scale
shifted 0.3 pixels bluewards and the G140M wavelength scale 0.3 pixels
redwards, and (3) the E230M wavelength scale shifted 0.3 pixels
redwards and the G140M scale 0.3 pixels bluewards.  Introducing even
these conservative relative shifts of 0.6 pixels between the STIS
E230M and G140M spectra does have a significant effect on the
absorption line parameters, particularly those of the putative
\di. This is because the redshift of the main \hi\ component is
determined by the higher order Lyman series lines in the G140M
spectrum.  Introducing a shift between the G140M and E230M spectra
means that the position of the main \hi\ absorption in the \lya\ line
changes.  To compensate for this change in position, the parameters of
the putative \di\ and the red \hi\ component are also changed.  We
describe the effect of these shifts on the parameters of the putative
\di\ in section \ref{derr}.

\subsection{Contamination of the {\em z} = 0.701 system by systems at other redshifts}
\label{contam}

We attempt to identify all lines in the E230M spectrum to determine if
the $z=0.701$ absorption lines are contaminated by \hi\ or metal lines
from systems at different redshifts.  There are six absorption systems
with log[$N$(\hi)]$> 14.5$ present in the E230M spectrum.  They are at
redshifts of 1.0867, 1.0674, 1.0549, 1.0318, 0.7247 and 0.5272. The
\lyb\ line at $z=0.5272$ falls at $\sim1566.5$\AA, and may contribute
to the absorption seen at this wavelength in the G140M spectrum (see
Fig.~\ref{ll}). We did not include the region with this absorption
when we fitted the G140M spectrum. For each log[$N$(\hi)]$> 14.5$
system we searched the wavelength ranges covered by the G140M and
E230M spectra for strong associated metal lines that may contaminate
the $z=0.701$ absorption lines, including Si\,{\sc ii}, N\,{\sc iv},
O\,{\sc i} and C\,{\sc ii}.  Only two systems show any strong metal
lines: the $z=1.0867$ system has associated \si3\ absorption, and the
$z=1.0318$ system shows \nv\ absorption. Galactic Fe\,{\sc ii} and
Mn\,{\sc ii} absorption is also present.  We found no other strong
metal lines or higher order Lyman \hi\ transitions that may be blended
with the $z=0.701$ lines and affect our analysis.

\subsection{{\em z} = 0.602 absorption system} 

\begin{figure}
\label{z6}
\begin{center}
\includegraphics[width=0.49\textwidth]{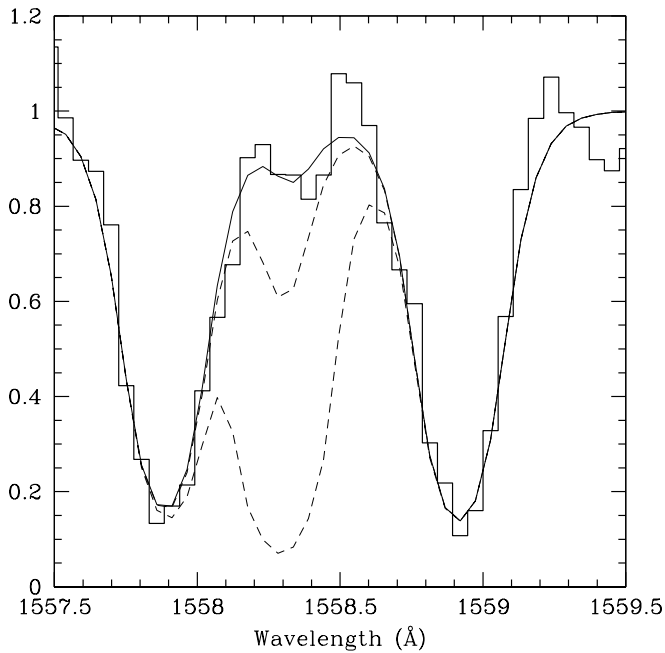}

\caption{A section of the STIS G140M spectrum showing the constraints
on the column density of the $z=0.602$ absorption system.  The thin
solid curve is the best fit obtained by including all \hi\ components
from the $z=0.70108$ system and a single component at $z=0.6023$. The
middle dashed curve shows the $z=0.6023$ column density set at the $2
\sigma$ upper limit. The lower dashed curve shows the column density
set to log[$N$(\hi)]$=16.7$, given in \citet{Tytler99}.  The $b$
parameter for the $z=0.6023$ line cannot be well determined by the
data and was fixed at $15$ \kms.  Increasing the $b$ parameter
decreases the upper limit on the $z=0.6023$ column density. }
\end{center}
\end{figure}

\citet{Tytler99} suggest that there is evidence in the IUE spectra for
another Lyman limit system with log[$N$(\hi)] $\simeq 16.7$ at
$z=0.602$.  It is important to determine whether such a system is
present, since if it is, its Ly-$\gamma$ line will fall at $\sim 1558$
\AA\ and may blend with the Ly-13 or Ly-14 lines of the $z=0.701$
system.  Such a Lyman limit system should have a \lya\ line at $\sim
1947$ \AA.  The STIS E230M spectra covers this region, although the
signal to noise is poor ($\sim 3$).  A strong line does appear at
$1948$ \AA, but the low S/N and contamination from a higher redshift
\hi\ \lyb\ line mean the column density cannot be determined from this
line alone.

However, we can put an upper limit on the column density of the
suspected $z=0.602$ system in the following way.  We fitted the
$z=0.701$ Lyman limit and higher order lines, excluding the lines
possibly blended with the $z=0.602$ Ly-$\gamma$ line.  Then we
compared the predicted Ly-13 and Ly-14 lines with the absorption seen,
assuming any residual absorption in these two lines is due to the
$z=0.602$ Ly-$\gamma$ line.  Combining the constraints from the low
S/N \lya\ line and the method above, we find that the contaminating
system has a redshift of $0.6023\pm0.00003$, placing it between the
$z=0.701$ Ly-13 and Ly-14 lines with log[$N$(\hi)]$ < 14.54$ (99.7\%
confidence).  The $z=0.602$ system does not affect the measured
parameters of the $z=0.701$ system, as illustrated in Fig.~\ref{z6}.
No metal lines were detected at $z=0.602$.

\subsection{Voigt profile fitting}
\label{voigt}

\begin{table*}
 \centering \begin{minipage}{140mm}
\begin{center} 
\caption{\label{tab} The absorption line parameters and their errors
given by VPFIT.  For the two \hi\ and putative \di\ lines these are
the best fitting parameters when the STIS Lyman limit abd the GHRS and
STIS \lya\ lines are fit simultaneously.  For \si3\ these are the best
fitting parameters when the GHRS and STIS \si3\ lines are fitted
simultaneously. The first errors given are the quadrature addition of
the statistical and continuum fitting errors.  The second errors for
the \si3\ line, \c4\ doublet and main \hi\ component are the errors
due to uncertainty in the wavelength calibration of the STIS spectra
(0.3 pixels).  Note that the errors for the putative \di\ and \hi\ red
sub-component are not meaningful (see section \ref{derr}). We list the
\mg2\ parameters published by KTOB based on observations taken with
Keck/HIRES. The error for $b$(\mg2) is our estimate of the error based
on our simulated spectra (see section \ref{voigt}).  The error for
$b$(\mg2) in KTOB is 1.1 \kms.}

\begin{tabular}{r l l l}

%RFC Second (logN) and last column (b) sig. figs changed 
\hline 
Ion & \multicolumn{1}{c}  {log($N$) (\cm2)} & \multicolumn{1}{c} %
z & \multicolumn{1}{c} {$b$ (\kms)} \\
\hline 
 \hi   & 17.213 $\pm$ 0.007 &0.701074 $\pm$ 0.000003 $\pm$ 0.000015& 22.1 $\pm$ 0.4 \\
 \hi\ \footnote{\small VPFIT does not return sensible errors for this \hi\ component. The reasons for this are described in section \ref{derr}.} & 13.5 &0.70142 & 19   \\
 Putative \di\ \footnote{\small See section \ref{derr}, Fig. \ref{nvz}, \ref {bvz} and \ref{bvn} for the putative \di\ parameter errors.} & 13.88 &0.701108 & 22 \ \\
 \si3 & 12.84 $\pm$ 0.05 & 0.701071 $\pm$ 0.000007 $\pm$ 0.000009& 14 $\pm$ 2 \\
 \c4 & 13.18 $\pm$ 0.08 & 0.701051 $\pm$ 0.000019 $\pm$ 0.000009& 20 $\pm$ 5 \\
 \mg2 & 11.50 $\pm$ 0.05 & 0.701100 $\pm$ 0.000005  & 12.0 $\pm$ 1.4  \\
\hline 
\end{tabular}
\end{center}
\end{minipage}
\end{table*}

Voigt profiles were fitted to the absorption lines and the best
fitting parameters were determined by minimising the $\chi^2$
statistic.  The program VPFIT\footnote{Carswell et al.,
http//www.ast.cam.ac.uk/$\sim$rfc/vpfit.html} was used for all line
profile fitting.  The absorption lines that were detected and fitted
in the $z=0.701$ system in the STIS spectra are the \lya\ line, the
Lyman series from Ly-$\epsilon$ to the Lyman limit, \si3\ ($1206.5$
\AA) and the \c4\ doublet ($1548.2$ and $1550.8$ \AA). The GHRS
spectrum covers the \lya\ and \si3\ lines.  We fitted all parameters
for case (1) described in section \ref{shift}, for no shift between
the STIS spectra.  The 0.3 pixel error in the uncertainty in the STIS
wavelength calibrations is listed after the statistical error in Table
1 for the lines it applies to.  We look at the effect of the shifts
between the STIS spectra wavelength calibrations on the putative \di's
parameters in section \ref{derr}.

The parameters of the main \hi\ component are tightly constrained by
the unsaturated \hi\ higher order lines and the drop in flux at the
grey Lyman limit.  The main uncertainties in this case are the
position of the continuum and the accuracy of the STIS wavelength
calibration.  Table \ref{tab} shows the best fitted parameters to the
main \hi\ component.  The error in redshift due to the error in STIS
zero point wavelength calibration is given after the statistical
error.  The first error for each parameter given in the table is the
quadrature addition of the statistical fitting error and the continuum
placement error, where applicable.

The \si3\ line parameters were found by fitting the GHRS and STIS
spectra simultaneously.  Again we include separately the statistical
error in redshift and the error when the wavelength calibration
uncertainty of the STIS E230M spectrum is taken into account.  The
\si3\ line is offset bluewards from the main \hi\ system by $0.5 \pm
1\textrm{ {\small (statistical)}}$ \kms.

The \c4\ line parameters were found by fitting the \c4\ doublet in the
STIS E230M spectra.  The statistical error and error due to wavelength
calibration are shown separately. The \c4\ lines are offset bluewards
from the main \hi\ component by $4 \pm 3 \textrm{ {\small
(statistical)}}$ \kms.  The best fitting $b$ parameter for \c4\ is
large compared to $b$(\mg2) and $b$(\si3). This difference is also
seen in many damped \lya\ absorbers \citep*{Wolfe00a} and is probably
due to \c4\ having a much higher ionization energy than \si3\ and
\mg2, and so being associated with a different velocity space in the
absorption cloud.

We use the parameters of the \mg2\ lines given in KTOB. To verify the
errors given by KTOB, we generated a synthetic \mg2\ spectrum from the
fitted $b$, $N$ and $z$ values they provided, at the same S/N
($\sim$70) and resolution (8 \kms\ FWHM) as the \mg2\ spectrum they
published.  Our $\sigma(N)$ and $\sigma(z)$ were very similar to
KTOB's, but our $\sigma(b)$ was larger, 1.4 \kms\ instead of 1.1
\kms. We use our larger $\sigma(b)$ estimate in our analysis. The
\mg2\ lines are offset redwards from the redshift of the main \hi\
system by $5 \pm 1\textrm{ {\small (statistical)}}$ \kms.  The
systematic error for the wavelength calibration for Keck/HIRES spectra
is $< 0.5$ \kms.  This $5$ \kms\ shift may be due in part to an error
in the STIS wavelength calibration (the 0.3 pixel error in wavelength
calibration we consider corresponds to a velocity error of $\sim$3
\kms).  Even though we are not sure of the cause of this shift, it does
not affect our analysis, since we use only the \mg2\ $b$ parameter
(section \ref{db}).

We attempted to fit the LL and both \lya\ lines with a single \hi\
component.  The minimum reduced $\chi^2 \equiv
\chi^2_{\rm{min}}/\textrm{(no. of degrees of freedom)}$ for this fit
was 1.7.  This fit predicted too much absorption in the higher order
lines, and could not account for extra absorption in the red and blue
wings of the \lya\ line.  We needed to add two extra absorption lines,
one \hi\ sub-component to the red of the main component, and a putative
\di\ line to the blue, to find an acceptable reduced $\chi^2$
($1.04$). Table \ref{tab} gives the parameters of the red and blue
\hi\ components.

\subsection{The errors on the parameters for the putative \di}
\label{derr}
\begin{figure*}
\centering
\begin{minipage}{140mm}
\begin{center}
\includegraphics[angle=270,width=0.99\textwidth]{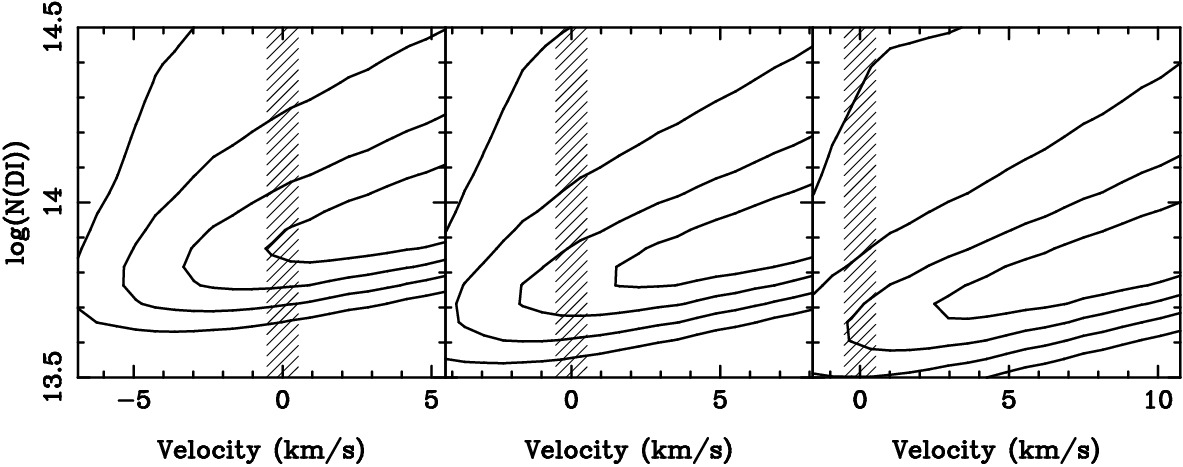}
\caption{\label{nvz} log$_{\rm{10}}$($N$[\hi]) versus velocity for the putative
\di.  For the left graph the Lyman limit is shifted 0.3 pixels
redwards and the \lya\ lines are shifted 0.3 pixels bluewards. For the
right graph the Lyman limit is shifted 0.3 pixels bluewards and the
\lya\ lines are shifted 0.3 pixels redwards. For the centre graph the
Lyman limit and \lya\ lines are fitted with no shift.  The velocity zero
point corresponds to 81.6 \kms\ blueward of the best fitting position
of the main \hi\ component, where we expect \di\ to be. The contours
represent, from innermost to outermost, the 68.4\%, 95.4\%, 99.73\%
and 99.99\% confidence levels. The vertical hashed region represents
the 68.4\% confidence region on the positon of the main \hi\
component.}
\end{center}
\end{minipage}
\end{figure*}

\begin{figure*}
\centering
\begin{minipage}{140mm}
\begin{center}
\includegraphics[angle=270,width=0.99\textwidth]{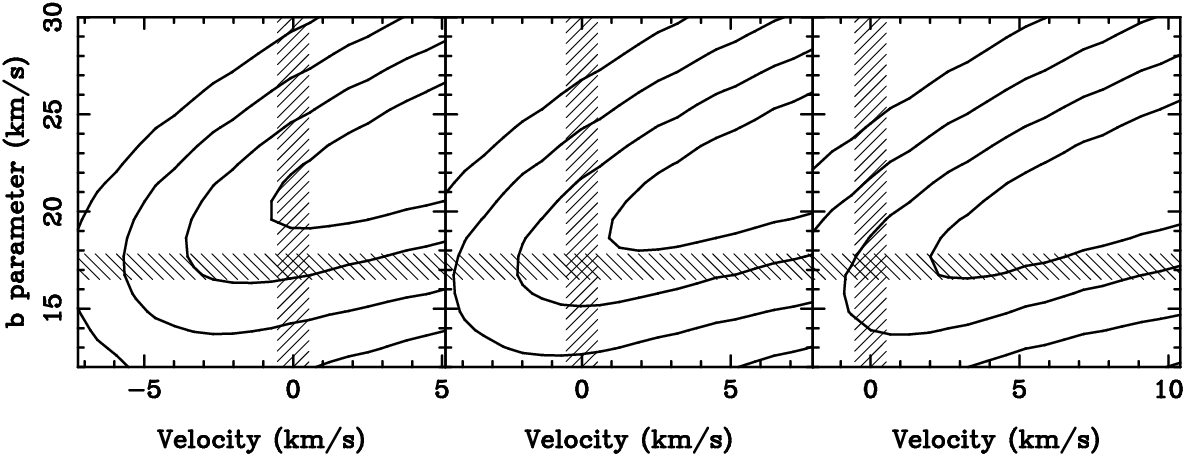}
\caption{\label{bvz} $b$ parameter versus velocity for the putative
\di.  For the left graph the Lyman limit is shifted 0.1 pixels
redwards and the \lya\ lines are shifted 0.3 pixels bluewards. For the
right graph the Lyman limit is shifted 0.3 pixels bluewards and the
\lya\ lines are shifted 0.3 pixels redwards. For the centre graph the
Lyman limit and \lya\ lines are fitted with no shift.  The velocity
zero point corresponds to 81.6 \kms\ blueward of the best fitting
position of the main \hi\ component, where we expect \di\ to be. The
contours represent, from innermost to outermost, the 68.4\%, 95.4\%,
99.73\% and 99.99\% confidence levels. The vertical hashed region
represents the 68.4\% confidence region on the positon of the main
\hi\ component. The horizontal hashed region represents the 68.4\%
confidence region for the \di\ $b$ parameter predicted by the \hi,
\si3\ and \mg2 (see Fig.~\ref{btest}). }
\end{center}
\end{minipage}
\end{figure*}

\begin{figure*}
\centering
\begin{minipage}{140mm}
\begin{center}
\includegraphics[angle=270,width=0.99\textwidth]{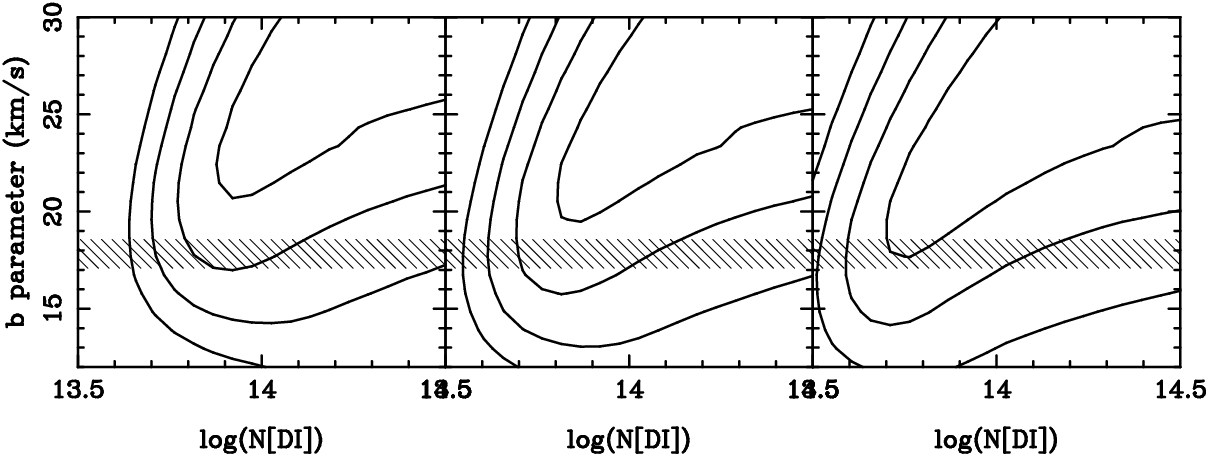}
\caption{\label{bvn} $b$ parameter versus log$_{\rm{10}}$($N$[\hi])
for the putative \di.  For the left graph the Lyman limit is shifted
0.3 pixels redwards and the \lya\ lines are shifted 0.3 pixels
bluewards. For the right graph the Lyman limit is shifted 0.3 pixels
bluewards and the \lya\ lines are shifted 0.3 pixels redwards. For the
centre graph the Lyman limit and \lya\ lines are fitted with no shift.
The contours represent, from innermost to outermost, the 68.4\%,
95.4\%, 99.73\% and 99.99\% confidence levels.  The horizontal hashed
region represents the 68.4\% confidence region for the \di\ $b$
parameter predicted by the $b$ parameters of \hi, \si3\ and \mg2 (see
Fig~\ref{btest}). }
\end{center}
\end{minipage}
\end{figure*}

The 1$\sigma$ errors on the parameters in Table 1 are those calculated
by VPFIT.  These errors are based on the assumptions that (1): the
model we have chosen (a single main \hi\ component with a red and blue
sub-component on either side) is correct and (2): the fitted
parameters are independent and the shape of the $\chi^2$ parameter
space around the minumum $\chi^2$ is parabolic. However, for the
putative \di\ the column density, $b$ parameter and redshift are
correlated. In addition, the $\chi^2$ parameter space is asymmetric
(see Fig.~\ref{nvz}, \ref{bvz} and \ref{bvn}).  This is due to the
small number of data points that are available to constrain the
parameters of the putative \di.  It may also be due to the difference
in `shape' between the STIS and GHRS \lya\ lines (see Figure
\ref{lya}).  For these reasons, the $1 \sigma$ estimates given by
VPFIT for the putative \di\ parameters may not be accurate.  This is
also likely to be the case for the red \hi\ sub-component.  For the
other lines (the main \hi\ component, \si3, \c4, and \mg2), the $1
\sigma$ errors given by VPFIT will be sufficiently accurate.

We find a robust parameter error estimate for the \di\ parameters by
using $\Delta\chi^2$ probability contour maps.  These are generated by
fixing two parameters (for instance, the $b$ parameter and redshift of
the putative \di) and then varying the remaining parameters (the
column density of the putative \di\ and all the parameters of the main
\hi\ component and red \hi\ component) to minimise $\chi^2$. This
creates a grid of $\chi^2$ values.  The distribution of $\Delta\chi^2
\equiv \chi^2 - \chi^2_{\rm{min}}$ is the same as that of a $\chi^2$
distribution with a number of degrees of freedom equal to the number
of fixed parameters (in this case, two). Here $\chi^2_{\rm{min}}$ is
the smallest value of $\chi^2$ for a particular grid.  Provided that
errors on the data points are normally distributed, the probability
contours generated in this way will be correct, irrespective of
correlations between the fitted parameters \citep*{Cash76}.  We also
need to consider the effect the shifts between the STIS E230M spectra
have on the \di\ parameters.  To do this we again consider the three
shifts described in the previous section.  Fig.~\ref{nvz}, \ref{bvz}
and \ref{bvn} show the probability contour maps for the three
parameters of the putative \di\ for the three different relative
shifts between the STIS Lyman limit and \lya\ spectra, representing
the uncertainty in the wavelength calibration between the two spectra.

\subsection{The putative \di's {\em b} parameter}
\label{db}

\begin{figure}
\begin{center}
\includegraphics[width=0.49\textwidth]{bplot.epsi}
\caption{\label{btest} The $b$ parameter squared (\kms)$^2$ versus the
inverse ion mass (amu$^{-1}$).  The dashed line is the least squares
line of best fit to the \mg2, \si3, and \hi\ points.  \c4\ and the
putative \di\ are shown for comparison.  The error bars in each case
are calculated from the 1$\sigma$ errors given by VPFIT.  Note that
the \di\ error bars are not particularly meaningful (see section \ref
{derr}).  The contour plots shown in Fig~\ref{nvz}, \ref{bvz} and
\ref{bvn} are required to determine whether $b$(\di) is consistent
with $b$(\hi), $b$(\mg2) and $b$(\si3).}

\end{center}
\end{figure}

We can predict what the $b$ parameter of a \di\ line associated with
the main \hi\ absorption from the $b$ parameters of the \si3, \mg2\
and main \hi\ component.  This is done by measuring the thermal line
broadening, $b_{\rm{therm}}$, and turbulent broadening,
$b_{\rm{turb}}$, by using the $b$ parameters from all available ions.
If we assume the absorbing cloud has a thermal Maxwell-Boltzmann
distribution and any turbulence can be described by a Gaussian
velocity distribution, then the $b$ parameter for a particular ion
will be given by

\begin{equation}
b_{\rm{ion}}^2 = b_{\rm{therm}}^2 + b_{\rm{turb}}^2.
\end{equation}

Here $b_{\rm{therm}}^2 = \frac{2kT}{m}$, where $T$ is the temperature
of the gas cloud, $m$ is the mass of the absorbing ion and $k$ is
Boltzmann's constant.  $b_{\rm{turb}}^2$ represents the Gaussian
broadening due to small scale turbulence, and is the same for all
ionic species. $b_{\rm{therm}}^2$ is proportional to the inverse of
the ion mass.  We plot $b^2$ against inverse ion mass in
Fig. \ref{btest}.  Subject to the assumptions above, all the ions in
the same cloud velocity space should lie on a straight line whose
intercept gives $b_{\rm{turb}}$ and slope gives the cloud temperature.
Since \mg2\ and \hi\ have very similar ionization potentials (IP)
(13.6 eV and 15.0 eV respectively), they should trace out the same
gas.  The IP of \si3\ is somewhat higher (33.5 eV), but there is
evidence that `intermediate' IP ions, such as Al\,{\sc iii}, have a
similar velocity structure to lower IP ions, such as \mg2\
\citep{Wolfe00a}.  We assume that the \mg2\ and \si3\ lines are in the
same gas as the \hi\, and fit least squares line of best fit to these
three points.  This is shown as the dashed line in Fig.~\ref{btest}.
If we use KTOB's estimate for \mg2\ $\sigma(b)$ . The \di\ point is
plotted with its $1 \sigma$ error bars as given by VPFIT.  We find a
temperature of $2.05 \pm 0.12 \times 10^4$K and a turbulent broadening
of $12.1 \pm 1.3$\kms.  Although we use our estimate of \mg2\
$\sigma(b)$ rather than KTOB's, if we do use KTOB's \mg2\ $\sigma(b)$,
the derived temperature and turbulent broadening do not change
significantly.

\subsection{Is the blue \hi\ component deuterium?}
\label{isd}
If the blue component is \di, it should fall at a redshift
corresponding to a velocity difference of 81.6 \kms\ bluewards of the
main \hi\ component. In addition, it should have a $b$ parameter
consistent with that predicted in the last section.

We use the contour plots described in section \ref{derr}, combined
with the $b$ parameter test in the last section, to see if the
putative \di\ component's parameters are consistent with those of \di.
Fig.~\ref{nvz}, \ref{bvz} and \ref{bvn} show how $\Delta \chi^2$
changes for a range of $b$ parameter, redshift and column density
values. Note the degeneracy between the parameters, which has a
consequence that the errors obtained using VPFIT are not likely to be
accurate for the \di.  The hashed bands on each of the plots show the
68\% confidence region that the parameters of the blue component are
expected to fall in if it is \di.  Fig.~\ref{bvz} best illustrates the
situation --- if the blue component is \di, we would expect its
redshift to be the same as that of the main \hi\ component (0\kms\
velocity difference) and the $b$ parameter to be consistent with the
$b$ parameters of the \hi\ and metal lines ($\sim 17$ \kms). 

For the left case of Fig.~\ref{bvz}, corresponding to a fixed relative
shifts between the G140M and E230M spectra, we can see that the $b$
parameter is completely consistent with that of \di, but the redshift
is inconsistent. For the right case, the redshift is consistent, but
the $b$ parameter is not. Where there is no relative wavelength shift
between the spectra, the parameters of the blue component are
inconsistent at a $\sim 80$\% confidence level.  While these results
do not suggest that the blue component is \di, they do not
conclusively show that is is not \di.

If we take the case of no shift between the G140M and E230M spectra
and {\em assume} the blue absorption is entirely due to \di,
fixing the $b$ parameter and redshift at those we expect for \di, we
find the ratio D/H $= (3.0 - 4.6) \times 10^{-4}$ ($1 \sigma$ limit). 

\subsection{Comparison with previous analyses of this system}
\label{comp}

We have analysed this absorption system using the same spectra
presented in \citet{Webb97a}, \citet{Tytler99} and KTOB, considering
important systematic effects not addressed in these previous analyses.

There are several small differences in our fitted parameters for the
main \hi\ component, putative \di\ component and \si3\ line compared
to KTOB's fitted parameters.  Our line spread function is slightly
different from KTOB's and we use a larger error for $b$(\mg2) (see
section \ref{voigt}).  We have explored the effect that relative
shifts between the wavelength scales of the G140M and E230M STIS
spectra have on the parameters of the putative \di\ line.  However,
none of these effects substantially change the conclusions of KTOB.
As can be seen from Fig. \ref{bvz}, if a large relative shift between
the G140M and E230M wavelength scales is present, the probability that
the putative \di\ really is \di\ is decreased.

The most important reason for our different result from KTOB is a
difference in generating our probability contours in Fig. \ref{nvz},
\ref{bvz} and \ref{bvn}.  When generating our $\chi^2$ contours, we
varied all the parameters of all the components in the $z=0.7$
complex.  To generate the plot in KTOB's Fig.~7, when calculating the
minimum $\chi^2$ only the column density of the putative \di\ component
was varied.  The remaining putative \di\ parameters and all the
parameters of the main \hi\ component and the red \hi\ sub-component
were fixed by KTOB.  This means their contours do not accurately
represent the actual range of probabilities for the putative \di\ $b$
parameter and redshift.  The true confidence ranges are somewhat
larger.  This is the main reason we find that the putative \di\
parameters are inconsistent with those expected for \di\ at the $\sim
80\%$ rather than $>98\%$ level.

Our D/H upper limit calculated assuming the blue absorption is
entirely due to \di\ is, not surprisingly, consistent with the
\citet{Tytler99} upper limit of $5.7\times 10^{-4}$.

\section{Discussion}

We have analysed the $z = 0.701$ absorption complex towards PG
1718$+$4801 using HST STIS and GHRS spectra, considering systematic
errors that were neglected in previous analyses.  The most important
of these systematic errors is the absolute wavelength calibration of
the STIS G140M and E230M spectra.  Introducing a relative shift,
comparable to the wavelength error, between these two sets of spectra
significantly alters the parameters of the putative \di.  

We find that the parameters of the absorption line previously
identified as \di\ are marginally inconsistent with those expected.
In particular, the $b$-parameter and line position of this feature are
at best consistent with the expected values at the 20\% level.

We have shown that the absorption system at $z = 0.602$, previously
claimed as having log[$N$(\hi)] $\simeq 16.7$ in fact has
log[$N$(\hi)] $< 14.5$.

If we assume that the blue component is \di, we find D/H $= (3.0 -
4.2) \times 10^{-4}$. If significant contamination is present, then of
course D/H may be significantly lower than $3.0\times 10^{-4}$.

The D/H range given here corresponds to a $1 \sigma$ range for
$\Omega_b h^2$ of $0.003 - 0.006$.  We can compare this range with
$\Omega_b h^2$ estimates from other sources.  Two recent estimates of
the primordial Li$^7$/H abundance are $(0.91 - 1.91) \times 10^{-10}$
\citep*{Ryan00a} and $(1.92 - 2.49) \times 10^{-10}$ \citep*{Bonif02a}.
Combining these two ranges, we find Li$^7(\Omega_b h^2) = 0.0044 -
0.020$.  Two samples recently used to measure the He$^4$ abundance,
$Y_P$, give $0.238 \pm 0.003$ (The sample in \citealp*{Izotov98a}, using
the ionization correction factor in \citealp*{Gruen02a}) and $0.2405 \pm
0.0017$ \citep*{Peim00a}.  Taking the highest and lowest limits of
these two ranges gives He$^4(\Omega_b h^2) = 0.0065 - 0.0145$. Sievers
et al. (2002) give CMB$(\Omega_b h^2) = 0.023 \pm 0.003$ based on all
currently available CMB data.  Taking an average of the five D/H
measurements made in other QSO absorption clouds, weighted by their
inverse variances, gives D$(\Omega_b h^2) = 0.021 \pm 0.001$.

The $\Omega_b h^2$ range for the $z=0.701$ QSO absorber is consistent
with the Li$^7$ range and only marginally inconsistent with the He$^4$
range.  It is significantly inconsistent with both the CMB and low QSO
D/H $\Omega_b h^2$ ranges, however.  The metallicity of this absorber
is low, [Si/H] $= -2.4$ \citep{Tytler99}, and the QSO absorption
systems used to measure D/H have similarly low metallicities.  At
these metallicities no significant astration is thought to have
occurred and the measured D/H values should be primordial.  For a
standard homogenous BBN, we expect the primordial D/H to be
independent of direction and redshift. Thus the inconsistency of the
$z=0.701$ D/H value with lower D/H values in other QSOs and the CMB
$\Omega_b h^2$ value supports the argument that the putative \di\ in
this system is contaminated with \hi.

However there are several caveats.  Only five D/H measurements in QSO
absorbers have been made.  Even among these `low' D/H values there is
considerable scatter, which is not expected based on the standard BBN,
no astration picture.  This is may be due to unaccounted for
systematic errors, but other explantions have been suggested.
\citet*{Fields01} suggest the scatter may be evidence for an early
population of stars that have caused astration even in these low
metallicity absorbers.  \citet*{Jedam02} suggest it is possible that
there may be rare astrophysical sites where D/H has been enhanced
above the primordial value.  It would be imprudent to reject entirely
the possibility that this system shows a high D/H until further D/H
measurements in QSO absorbers are available, particularly if there is
a trend of D/H with metallicity.

We acknowledge helpful comments from Michael Murphy and Mike Irwin,
and correspondence regarding STIS wavelength calibration with Claus
Leitherer at the STScI. We also thank Gary Steigman for his helpful
correspondence, and the referee for his suggestions on how to improve
the paper.

\bibliographystyle{/home/nhmc/tex/mnras/mn2e}
\bibliography{/home/nhmc/tex/references}

\end{document}